\documentstyle[12pt]{article}
\newtheorem{theorem}{Theorem}[section]
\newtheorem{lemma}[theorem]{Lemma}

\newcommand{\PI}{\mbox{$P_{\rm I}$}}
\newcommand{\PS}{\mbox{$P_{\rm S}$}}
\newcommand{\SC}{\mbox{$S$}}
\newcommand{\WC}{\mbox{$W$}}

\newcommand{\bq}{\begin{equation}}
\newcommand{\eq}{\end{equation}}
\input{mssymb}
\begin{document}
\title{Chaotic, regular and unbounded behaviour in
the elastic impact oscillator}
\author{Harbir Lamba}
\footnotetext{\noindent School of Mathematics, University of Bristol,
University Walk,
Bristol BS8 1TW, UK.  e-mail: h.lamba@bristol.ac.uk}
\date{July 1993}
\maketitle

\begin{abstract}
\noindent A discontinuous area-preserving mapping derived from a
sinusoidally-forced impacting system is studied.
This system, the elastic
impact oscillator, is very closely
related to the accelerator models of particle physics such as the Fermi map.
The discontinuity in the mapping is
due to grazing which can have a surprisingly
large effect upon the phase space.
In particular, at the boundary of the stochastic sea, the discontinuity set
and its images can act as a partial barrier which
allows trajectories to move between
chaotic and regular regions.
The system at higher energies is also analysed and Moser's invariant curve
theorem is used to find
sufficient conditions for the existence of invariant curves that bound the
energy of the motion. Finally the behaviour of the system under more general
periodic forcing is briefly investigated.
\end{abstract}
\newpage
\section{Introduction}
\label{sec:intro}
We study a forced oscillator, such as a
mass on a spring, moving in one dimension and repeatedly
impacting against a fixed wall (Figure~\ref{fig:imposc}).
The motion between impacts is undamped and
the impact is modelled as
an instantaneous reversal of direction with a constant coefficient of
restitution. For a linear spring with
sinusoidal forcing this system, called the {\em impact
oscillator}, is described after suitable rescaling
by the following equations
\bq
\begin{array}{rcll}
\ddot{x} + x &=&\cos(\omega t), \;\;\;\; & x < \sigma \\
\dot{x} & \mapsto & -r \dot{x}, & x=\sigma
\label{eq:imposc}
\end{array}
\eq

The inelastic impact oscillator $(r<1)$,
was first studied \cite{sh83a,sh83b,tg82,ts86}
as a model for many important engineering phenomena including the rattling
of engine components and the behaviour of structures in
earthquakes. Further work has been done
on such aspects as the bifurcation structure and
the form of the chaotic attractors \cite{w87a,w87b,w92,d92,bd93}.
Here we consider the elastic case
$r=1$ where
no energy is lost at the impacts.
For engineering applications
$r$ is often close to 1
and the elastic limit is an efficient way of exploring large
areas of the phase space. It is also possible to exploit the
time-reversal symmetry of the elastic case to locate the
periodic orbits. However the elastic case
has unusual behaviour of its own, and that is the subject of this
paper.

The motion of the impact oscillator consists of a series of
impacts separated by smooth motion between the impacts.  It is
natural to describe these systems in terms of the
{\em impact map}, \PI, relating the state of the system at one impact
to the state at the next.
Each impact is described by the (positive) velocity of the mass just before
the impact and the
phase (from $0 \mbox{ to } 2 \pi$) of the forcing cycle at the
time of impact so that
the phase space is a half-cylinder.
Impacting systems that have been studied before include
the accelerator models of particle physics \cite{u61,b71,ll72,p78,llc80,ll,p87}
and billiards, see \cite{b81,ks86,hw83} and references therein.

Although a mechanical impacting system is clearly discontinuous when regarded
as
a continuous-time dynamical system,  the resulting impact map may be analytic
and indeed most of the literature concerns impacting systems with impact maps
that are either smooth or have discontinuities deliberately
introduced into a
higher derivative.
The map \PI\ is discontinuous because of grazes, or zero-velocity impacts,
see trajectory B of Figure~\ref{fig:graze},
where the mass approaches the wall, touches it, and is then pulled away
again. Nearby trajectories (trajectory C) will either (just) hit
the wall and be deflected by it or miss the wall and go on to impact at a
later time (trajectory A). Surprisingly,
low velocity impacts strongly distort the phase space and
so have a large effect upon the overall dynamics.
This process is
described in \S\ref{sec:sing} and is an important source of
chaotic behaviour. Previous work \cite{w92,d92,n91a,n91b,fb92}
has shown that grazing
can introduce additional bifurcations such as the sudden
appearance or disappearance of periodic orbits and the interruption of
period-doubling cascades.

The phase space of \PI\ resembles that of
the Fermi map \cite{ll72,llc80,ll} in that
the motion becomes more
regular as the velocity of the impacts increases. Figure~\ref{fig:pspace}
shows a phase space plot of \PI\ for
parameter values
$\sigma = 0.1, \, \omega = 2.5 $  which has clearly separated
regions of phase space giving rise to chaotic and regular behaviour.  In
particular, at low velocities the phase space is filled
by a stochastic
sea which, as the velocity increases, is interrupted by regular islands
associated with elliptic fixed and periodic points.
Above the stochastic sea the behaviour is more regular, with bounding
invariant curves and elliptic islands.
Note that the upper boundary of the stochastic sea is
the lowest bounding invariant curve.

However for slightly different values of $\sigma$ and $\omega$ we
observe a new phenomenon.
Figure~\ref{fig:mixed} shows
a {\em single\/}
trajectory for the parameter values $\sigma = 0$, \,
$\omega = 2.8$. The trajectory starts in the stochastic sea,
exhibiting a typical sensitivity to initial data,
but then, at irregular intervals, leaves the sea and moves along a smooth
curve before reentering the stochastic sea.
There is, for this example, less of a distinction between chaotic
and regular
motion as both can occur on the same trajectory. This very unusual behaviour
is due to the discontinuity introduced by the grazing.
The grazing discontinuity set and its iterates form structures at
the upper boundary of the stochastic sea which act as a partial
barrier allowing
trajectories to move between the chaotic and the regular regions.
Partial barriers also occur for smooth systems although the mechanisms are
quite
different
(see MacKay et al. \cite{mmp84}).
The existence of trajectories that can behave both regularly and chaotically
is of great importance to the understanding and
modelling of this system.
In an engineering context, it is also important that such trajectories
involve a large number of high-velocity impacts
(with associated high wear-rates) and is all
the more surprising since they arise directly from low velocity impacts.


The outline of the paper is as follows.
In \S\ref{sec:imposc} we define the impact oscillator and the impact map
and in \S\ref{sec:sing}
we analyse the effect of the grazing discontinuity on the
nearby dynamics.
In \S\ref{sec:barriers} we examine the upper boundary of the
stochastic sea and show how interactions between the
grazing discontinuity set and invariant curves
can allow trajectories to alternate
between stochastic and regular regions.
\S\ref{sec:discont}
describes the relationship between the elastic impact oscillator and other
impacting systems such as the accelerator models and billiards.
Then in \S\ref{sec:highv}, motivated by the links with the accelerator
models, we turn our attention to
the behaviour of the system at high
velocities where the impact map is smooth.
We show that the existence of bounding invariant curves has an interesting
dependence on the
parameter values
and briefly look at the case of more general
periodic forcing.

\section{The impact oscillator and the impact map}
\label{sec:imposc}

The impact oscillator is a mass on a linear spring, moving in one
dimension
subject to a sinusoidal forcing, which impacts against a rigid obstacle
referred to as the {\em wall\/} (Figure~\ref{fig:imposc}).
The impact itself is modelled as an
instantaneous process with coefficient of restitution $r$, $0 \leq r \leq 1$.
By rescaling the system so that the amplitude of the forcing and the natural
frequency of the spring are equal to 1
we obtain equation (\ref{eq:imposc})
where $\omega$ is the {\em forcing frequency\/}
and $\sigma$, the {\em clearance}, is the
position of the wall relative to the equilibrium position
of the mass in the unforced system. Thus the system has three
parameters $r, \sigma$ and $\omega$. In this paper $r=1$.

If the impacting condition is removed then the system is just the forced
harmonic oscillator --- a linear system whose behaviour is completely
understood. But the impact oscillator is highly nonlinear
and cannot be regarded as a small perturbation of a linear or integrable
system.
It is convenient
to replace the time, $t$, by the {\em phase}, $\phi$, which is
defined by $\phi = (\omega t) \bmod 2 \pi $.
So we have the state vector
${\bf u} = (\phi, x, v)$, where $v$ is the velocity $\dot{x}$,
and the phase space $\Omega = 2\pi S^{1} \times
(- \infty,\sigma] \times \Bbb R$. The vector ${\bf u}$ evolves under the
discontinuous flow $\Phi$.

A common approach to studying a system with periodic forcing is to define
a Poincar\'{e} surface
$\Sigma_{S} = \left\{(\phi,x,v):\phi=\phi_{0}\right\}$
(different values of $\phi_{0}$ give different maps
but the choice is unimportant). This defines the {\em stroboscopic map},
\PS, which maps the position and velocity at the phase
$\phi_{0}$ of the forcing
cycle onto the position and velocity one cycle later.
Note that $\Sigma_{S}$ is always crossed transversely since time is
increasing.
But it is the impacts themselves that are of interest
and the map \PS\  tells us very little about these
--- for example, during a single forcing cycle there may be no impacts
or there may be many and determining which is the case would appear to be a
very hard problem.

Instead we study the mapping \PI\ which maps one impact onto the
next one (\cite{sh83a,sh83b,w87a}).
\PI\ is an extremely useful tool for the analysis of the impact oscillator.
It contains all of the interesting dynamical information, since the
impacts are the source of the nonlinearity and the motion between
them is just that of a simple harmonic oscillator.

The map \PI\ is discontinuous due to {\em grazing
trajectories}. These are trajectories which pass through a point
$(\phi,\sigma,0)$, known as a {\em graze}. Nearby orbits either have a
low velocity impact close to $(\phi,\sigma,0)$ or miss the wall and impact
at a later time
and so \PI\ is discontinuous in the neighbourhood of grazes.

We now examine the map \PI\ and its discontinuities in more
detail.

\subsection{The impact map}

We define the Poincar\'{e} surface $\Sigma_{I} =
\left\{(\phi,x,v):x=\sigma,
v > 0\right\}$. The resulting map takes the phase and velocity
$(\phi_{n},v_{n})$ at one impact
to the phase and velocity
$(\phi_{n+1},v_{n+1})$ of the next impact. Conventionally, the
velocity is recorded just before the impact and we find it
convenient to {\em not\/} regard grazes as impacts. Thus the recorded velocity
is always positive. More formally,
the {\em impact map}, $P_{I}$, is defined by
\[\PI:\Sigma_{I} \rightarrow \Sigma_{I} \ {\rm where} \ \Sigma_{I}=2\pi S^{1}
\times \Bbb R^{+} \ {\rm and} \ \PI:(\phi_{n},v_{n})\mapsto
(\phi_{n+1},v_{n+1})\]
The phase space of \PI\ is therefore an open half-cylinder.
Technically, the impact map is not a Poincar\'{e} map since the
surface $\Sigma_{I}$ is not everywhere transverse to the flow because
of grazing trajectories.

The precise definition of \PI\ is actually more cumbersome than we have
described here. It is straightforward to show that since the
motion between impacts is recurrent
one impact must lead to another
\cite{w87b,d92} and so the map is defined everywhere.
However, if $|\sigma| < 1$, it is possible for the mass to
stick to the wall until the acceleration becomes negative and it moves away.
This behaviour occurs with measure zero and if it
occurs more than once on a single orbit, the motion must be periodic
(in contrast, for the inelastic case $r < 1$, the mass
can stick to the wall via an
infinite sequence of bounces which brings the mass to rest in finite time
and for some parameter ranges this behaviour can be extremely important).

The free motion (between impacts) is described by the first equation
of (\ref{eq:imposc}) which is just the forced harmonic oscillator and
can be solved exactly.
If $\omega \neq 1$ and the time and velocity of an impact are
given by $t_{0}  \ {\rm and} \ v_{0}$ then the position and velocity
of the mass at time $t$ are given by
\begin{eqnarray}
x(t;t_{0},v_{0})&=&(\sigma-\gamma \cos\omega t_{0}) \cos(t-t_{0})\nonumber \\
&&\mbox{} +(-v_{0}+\omega \gamma \sin\omega t_{0})\sin(t-t_{0})
+ \gamma \cos\omega t
\label{eq:pos}
\end{eqnarray}
\begin{eqnarray}
v(t;t_{0},v_{0})&=&(-v_{0}+\omega \gamma \sin\omega t_{0})
\cos(t-t_{0}) \nonumber \\
&& \mbox{} -(\sigma-\gamma \cos\omega t_{0})\sin(t-t_{0})
-\omega
\gamma \sin\omega t \label{eq:vel}
\end{eqnarray}
where $\gamma = 1/(1-\omega^{2})$. This solution is only
valid for $t_{0} \leq  t < t_{1}$ where $t_{1}$ is the time of the next
impact given by the first solution of the transcendental equation
\bq x(t_{1};t_{0},v_{0})=\sigma \label{eq:transc} \eq
To solve the system numerically we only have to find the time of the next
impact using a root-finding scheme. This is both quick and accurate since no
numerical integration is involved although there is always the
possibility of not detecting extremely low-velocity impacts.

Implicit differentiation of (\ref{eq:pos}), (\ref{eq:vel})
and (\ref{eq:transc}) gives an expression
for the Jacobian derivative of \PI\ (see \cite{sh83a})
\begin{eqnarray}
\lefteqn{D\PI (\phi_{0},v_{0})=}  \nonumber \\ \label{eq:DPI}
&& \left( \begin{array}{cc}
(N_{0} S -v_{0} C) / v_{1} &  S / v_{1} \\
&\mbox{} \\
( N_{0} N_{1} / v_{1} -  v_{0} ) S
- ( N_{0} + N_{1} v_{0} /  v_{1} ) C
&  N_{1}S/v_{1} - C
\end{array} \right)
\end{eqnarray}
where $S= \sin(t_{1}-t_{0}), \ C=\cos(t_{1}-t_{0})$ and $N_{i} = \cos(\omega
t_{i}) - \sigma , \ \ i=0,1$. The $N_{i}$ are the accelerations of the mass
just before the
impacts. Note that $t$ appears instead of $\phi$ on the right-hand
side of (\ref{eq:DPI}). This is purely for convenience.

It is immediate that
\[ \left| D\PI \right|  = \frac{v_{0}}{v_{1}}  \]
as stated in \cite{sh83a}. In fact this is equivalent to showing that
\PI\ preserves the measure $v\,d\phi\,dv$ and
we can simplify the
determinant by introducing the coordinate change $z=v^{2}$. Writing \PI\
as a map from $(\phi,z)$ to $(\phi,z)$ gives us the Jacobian derivative
\begin{eqnarray}
\lefteqn{D \PI (\phi_{0},z_{0}) =} \hspace{-.5in} \nonumber \\
&& \hspace{-0.5in} \left( \begin{array}{cc}
(N_{0} S-\sqrt{z_{0}}C)/\sqrt{z_{1}} & S/2\sqrt{z_{0}z_{1}} \\
\mbox{} & \mbox{} \\
2(N_{0}N_{1}-\sqrt{z_{0}z_{1}})S -2(N_{0}\sqrt{z_{1}}+N_{1}\sqrt{z_{0}})C
& (N_{1}S-\sqrt{z_{1}}C)/\sqrt{z_{0}} \end{array} \right)
\label{eq:newjac} \end{eqnarray}
which has determinant equal to 1 and so is an area-preserving mapping.

The elements of $D\PI$ depend on both $(\phi_{0},v_{0})$ and
$(\phi_{1},v_{1})$ and so the Jacobian is defined implicitly. Also the elements
of (\ref{eq:DPI}) become unbounded as
$v_{1} \rightarrow 0$, that is, as the next impact
tends to a graze. \PI\ is therefore a map from
the half-cylinder to itself which is everywhere smooth except on the
1-dimensional set \SC, the {\em discontinuity set}, where
\[ \SC = \lim_{\epsilon \rightarrow 0} \left\{(\phi_{0},v_{0}):\PI
(\phi_{0},v_{0})=(\phi,\epsilon) \mbox{ for some }
 \phi \in 2\pi S^{1} \right\}\]

The set \SC\ can be thought of as the pre-image of the line $v=0$.
The sets \SC\ are included in Figures~\ref{fig:pspace} and \ref{fig:mixed} and
a more complicated example is shown in Figure~\ref{fig:sc}.
\SC\ is not, as one might expect,  a smooth closed curve
that spans the phase space but instead consists of one or more smooth
curve segments.
This is because for part of the
line $v=0$ the acceleration is positive and the mass sticks to the wall.
Also there are trajectories that have consecutive grazes and these
cause the smooth segments of
\SC\ to connect to one another in a complicated way as shown in
Figure~\ref{fig:sc}.
For a detailed description of \SC\ see
\cite{w92}. An important result that we shall need in \S\ref{sec:barriers}
and \S\ref{sec:highv}
is that \SC\ is bounded above. This means that for sufficiently high velocities
the impact map will be smooth.

The time-reversal symmetry of the system
means that the phase space of \PI\ is invariant
under the transformation
\bq t \mapsto -t, \quad \phi \mapsto 2\pi-\phi \label{eq:symm} \eq
and is symmetric about the midline $\phi=\pi$.
The forward image of the line $v=0$ is therefore the image of
\SC\ under the  time reversal symmetry (\ref{eq:symm}). We call this line \WC.
We define further iterates and preiterates of these sets by
\[\SC^{n} = \PI^{1-n}(\SC), \quad \WC^{n} = \PI^{n-1}(\WC)\]

The map \PI\ behaves very differently on each side of \SC.
On one side of the curve the
trajectories have low velocity impacts and the phase space becomes very
distorted as can be seen from the singular Jacobian for $v=0$.
On the other side, the trajectory just misses the wall and hits it at
a later time and the Jacobian does not have large elements.

\section{The dynamics of the impact map near $S$}
\label{sec:sing}
To illustrate the nature of \PI\ close to the discontinuity set \SC\
we imagine a line segment $I$ of initial conditions in phase space which
transversely crosses \SC\ (Figure~\ref{fig:singset}). Let $I$ and $S$ meet at
point $B$ and call
the resulting two sections of $I$, $I^{+}$ and $I^{-}$. The particle motions
corresponding to the points $A,B \mbox{ and } C$ are those shown in
Figure~\ref{fig:graze}.
As we move along $I^{+}$ from $A$ to $B$ the image curve $\PI (I^{+})$ is
traced out and moves towards the line $W$ and meets it
transversely. The endpoint
$\PI (B)$ lies on the line $W$ but is not (generically) the image of the point
$B$ under the symmetry transformation.
As we continue from $B$ to $C$
the next impact is now
a low velocity impact. So the curve $\PI (I^{-})$
grows out of the line $v=0$ and the line $I$ would appear to be split in two.
However, the second iterate of $I^{-}$ rejoins $\PI (I^{+})$ meeting it at
$\PI (B)$ and
Whiston \cite{w92} showed that $P^{2}_{I}(I^{-})$ meets \WC\
tangentially. The side of \SC\ that does not map directly to low
velocity impacts is called the {\em non-grazing side\/} and the side
that does is the {\em grazing side}.
The line $\PI^{2}(I^{-})$ is locally stretched by a factor of
of ${\cal O}(\epsilon^{-\frac{1}{2}})$ where $\epsilon$ is the
distance from \SC.
This stretching is described by a square-root singularity.
Together the cutting and stretching
have important implications for the dynamics. Because grazes tend to collapse
trajectories onto \WC, \WC\ and its iterates strongly influence the
overall dynamics.
The stretching also means that periodic orbits which include
low velocity impacts,
$v \ll 1$,
are highly likely to be unstable and so lie in the stochastic sea
as can be easily seen by examining the Floquet
multipliers derived from (\ref{eq:DPI}) or (\ref{eq:newjac}).

We end this section with a few final comments about the map \PI.
The impact oscillator system is unusual in that the preferred Poincar\'{e}
section permits tangential intersections with the flow.
This is not because there is no
available everywhere-transverse section ($\Sigma_{S}$ for example)
but because the impact map is a much more powerful tool for studying
the impacts which are the source of chaos in the system.
It is the tangential intersections with the Poincar\'{e} surface that
define the sets \SC\ and \WC, but it is the low velocity impacts that
are responsible for the cutting and stretching that is observed near \WC\
Because $x=\sigma$ is both the Poincar\'{e} surface
and the impacting surface these effects are seen together.

\section{Chaotic motion and partial barriers}
\label{sec:barriers}

We now study the main chaotic region, often
known as the stochastic sea. The two
phase space plots, Figures~\ref{fig:pspace} and \ref{fig:mixed},
show the stochastic sea lying between the
line $v=0$ and the regular region which exists at high velocities.
It is our
intention to explain the novel behaviour seen in Figure~\ref{fig:mixed} where
trajectories move between the stochastic sea and the regular region lying
above it. To do this we must look carefully at the interaction
between the discontinuity set and regular curves
at the upper boundary of the
stochastic sea.

\subsection{The discontinuity set and the stochastic sea}

As we have already shown, the set \SC\ is a very strong source of chaos in the
system due to the arbitrarily large stretching close to a graze. From
numerical experiments it appears to be impossible
for an invariant curve to cross \SC\ without being destroyed.
This agrees with previous studies \cite{b71,hw83}
which observed that invariant
curves do not usually survive crossing
lines of discontinuities in the first or second derivative
and implies that
\SC\ can only lie within chaotic regions. It is easy to obtain
an upper bound on \SC\ from equations (\ref{eq:pos}) and (\ref{eq:vel}) and
also
straightforward to show that \SC\ must touch the line $v=0$ (for all
$r, \sigma$ and rational $\omega$
there exists a smooth motion which never hits the wall but
repeatedly grazes it). These facts strongly suggest (but do not prove) that the
set \SC\ will lie in a single bounded
chaotic region extending upwards from $v=0$,
namely the stochastic sea. This
is supported by the numerical studies.

\subsection{The boundary of the stochastic sea}

First let us consider the case where \SC\ lies well away from the boundary of
the sea.
This is illustrated
by Figure~\ref{fig:pspace} where \SC\
lies well inside the sea. The boundary is the lowest
invariant curve that spans the cylindrical phase space and since the map is
smooth in the neighbourhood of this curve the boundary is the same as those
found in smooth area-preserving maps, with regular curves above it and a
sticky chaotic layer just below it.

We now examine Figure~\ref{fig:mixed}.
This shows a single
trajectory, with initial condition lying within the
stochastic sea and followed for 20000 impacts. In
that time it makes 3 excursions (each containing many hundreds
of impacts)
onto regular curves surrounding two different period-5 orbits.
There is no smooth
boundary to the stochastic sea,
but a {\em partial barrier}
which allows trajectories to move between the stochastic
sea and the elliptic islands where the trajectory moves slowly around
regular curves until it reenters the sea.
This is because for these parameter values \SC\
does not lie within the stochastic sea, but
touches the boundary
and intersects the islands associated with the
two elliptic period-5 orbits.

In order to understand this mechanism
we first look at how a trajectory moves from an elliptic curve into the
stochastic sea.
Figure~\ref{fig:blowup} shows a schematic blowup of the last few iterates of
the motion along the elliptic curve that is intersected by \SC\ (so only every
$5^{{\rm th}}$ iterate is shown).
The trajectory moves slowly down the curve until it
crosses \SC. The trajectory is now on the grazing side of \SC\ and so the next
impact (which is not shown) will be a low velocity impact which did not occur
on the previous cycle. The trajectory is now cut, stretched
and reconnected as described in
\S\ref{sec:discont} so that it returns to the neighbourhood of the
elliptic curve having been stretched tangentially along the line $\WC^{5}$ and
then very rapidly disappears into the stochastic sea.

The opposite mechanism, where the trajectory suddenly
jumps out of the stochastic sea
onto the regular curves is of course
just the time reversal of the above mechanism. So the corresponding diagram
is Figure~\ref{fig:blowup} with the order of the points reversed, \SC\
replaced by \WC\ and $\WC^{5}$ replaced by $\SC^{5}$.
It is not necessary for the system to be time-reversible in order
for trajectories to be able to leave the sea,
the existence of such trajectories
can be deduced
from the area-preserving property of the map --- if trajectories can enter a
region then trajectories must also leave that region.

We refer to the boundary in this case as a
partial barrier since the discontinuity set and its iterates form
structures which severely limit the rate of
transport (of phase space area)
across them but do not stop it entirely. Partial barriers for
smooth systems (see \cite{mmp84}) include cantori and the turnstiles associated
with periodic orbits. The mechanism described here is completely different
and, unlike these other cases, the motion on one side of the barrier is
regular.
Some sections of the partial barrier allow trajectories to move upwards while
others allow trajectories to move down. A section of the barrier is
enlarged in Figure~\ref{fig:turn}. For this piece of the barrier
the upward and downward
sections are separated by the invariant curve which just touches \SC\ and
surrounds the higher of the
period-5 elliptic orbits.

The lower of the two period-5 elliptic orbits (the one that appears to lie in
the stochastic sea) does not actually exist, even though part of the
surrounding curves do. It corresponds to
an unphysical motion --- that is, one in which the mass
moves through the wall instead of experiencing a
low-velocity impact. The position of these
unphysical orbits can
easily be found by modifying the numerical code so that low-velocity impacts
are ignored and the mass is allowed to move freely for a short time in the
forbidden region $x > \sigma$.

The precise nature of the partial barrier depends upon the number, type and
orientation of the regular curves and regions that are intersected by
\SC\ and it is possible to find parameter values for which the pictures can
become extremely complex, for example when \SC\ intersects higher order chains
of Birkhoff elliptic points.

There is however another simple example that is very important, shown in
Figure~\ref{fig:case2}, and that is when \SC\ intersects a region filled with
bounding
invariant curves rather than a elliptic region.
At first glance
Figure~\ref{fig:case2} looks just like
Figure~\ref{fig:pspace} with a stochastic sea
bordered by a smooth invariant curve.
However \SC\ now touches the boundary and the difference lies in the
position and the nature of this invariant curve.
It is the lowest bounding curve not because it is on the
point of breaking up, with regular curves above it and a sticky chaotic
border below, but because it lies immediately above \SC\ and all
the invariant curves below it have been destroyed.
Let us now consider the region just below this
bounding curve. Here the phase space consists
of invariant curves interrupted by \SC. Once a trajectory lands on one of
these curves (having been stretched along \WC\ by a low velocity impact) it
will
move regularly along that curve until it lands on that small part
which lies on the grazing side of \SC. It then
has a low velocity impact and moves back down into the sea.
This process is essentially the same as that
of Figure~\ref{fig:mixed}
--- the only difference being that
in Figure~\ref{fig:case2}
the regular curves interrupted by \SC\ are bounding curves
rather than elliptic curves and it is only the latter
which rise above
\SC\ and produce startling pictures such as
Figure~\ref{fig:mixed}.
So, while the stochastic sea in Figure~\ref{fig:case2} looks like a perfectly
ordinary chaotic region from a smooth system it does in fact display the same
alternating regular/chaotic behaviour as Figure~\ref{fig:mixed}.

The grazing mechanism provides an interesting new
transition between the regular
curves that exist at high velocities and the truly chaotic motion at very low
velocities.
If we replace the wall by a steep potential gradient then the
resulting smooth system should also display similar behaviour.

We end by noting that exactly the same
interactions between the discontinuity set and
regular curves that occur at the upper boundary
can occur at the boundaries between the chaotic region and
elliptic islands that lie within it.

\section{Related impacting systems}
\label{sec:discont}

Impacting systems and discontinuous
mappings have been studied before. To help put our work
into perspective we briefly look at two closely related classes of impacting
system which have been fundamental in the study of both smooth and
discontinuous
dynamical systems.

\subsection{Billiards}

Billiards are an important class of impacting system
\cite{b81,ks86,hw83}.
A billiard
is a point mass moving in a bounded
2-dimensional region with boundary ${\cal D}$.
When the mass hits ${\cal D}$ it bounces away elastically, following the
usual `angle-of-incidence equals angle-of-reflection' law.
A boundary component is
called {\em concave\/} or dispersing if it curves away from the bounded
region
and
{\em convex\/} or focussing if it bends inwards. A billiard is called
convex if the
bounded region is convex.
Natural, area-preserving, coordinates to use for the impact map are the
curvilinear distance,
$\eta$, along the boundary from some arbitrary point
and $s=\sin \alpha$ where $\alpha$ is the oriented angle between the
normal to the boundary and the incoming trajectory (see Figure \ref{fig:bill}).
The mapping from one impact $(\eta_{0},s_{0})$ to the next
$(\eta_{1}, s_{1})$
has an implicitly defined Jacobian \cite{ks86,hw83}
\begin{eqnarray*}
\lefteqn{J=\frac{\partial(\eta_{1} ,s_{1})}{\partial(\eta_{0},s_{0})}=} \\
&& \left( \begin{array}{cc}
\displaystyle \frac{C_{0}d-\cos \alpha_{0}}{\cos \alpha_{1}} & \displaystyle
-\frac{d}
{\cos \alpha_{0} \cos \alpha_{1}}
\\
&\mbox{} \\
\displaystyle C_{0} \cos \alpha_{1} + C_{1} \cos \alpha_{0} - C_{0}C_{1}d &
\displaystyle \frac{C_{1}d-\cos \alpha_{1}}{\cos
\alpha_{0}} \\
\end{array} \right)
\end{eqnarray*}
where $C_{0},C_{1}$ denote the curvature at the points of impact
and $d$ is
the Euclidean distance between the points of impact.
The form of this Jacobian in the limit $\cos \alpha_{1} \ll 1$,
which corresponds to the mass hitting a non-convex piece
of the boundary almost tangentially, is the same as
that of the impact
oscillator with area-preserving coordinates (\ref{eq:newjac}) in the
grazing limit
$ z_{1} \ll 1$.
In other words, a low velocity impact in the impact oscillator
corresponds to a nearly tangential impact in a non-convex billiard
which will have a discontinuity set at the pre-images of such
impacts.
In fact the cutting and stretching close to grazing trajectories
described in
\S\ref{sec:sing} and Figure~\ref{fig:singset}
is exactly the same for these non-convex billiards.

The most frequently studied billiards are those with
boundaries that are either everywhere-focussing or
everywhere-dispersing. Billiards that have boundaries with focussing and
dispersing components are likely to have both regular and chaotic regions
and the mechanism described in \S\ref{sec:barriers}
will also be of importance to the study of these systems.

\subsection{Accelerator models}

In 1949 Fermi \cite{f49} proposed a
mechanism for the acceleration of cosmic rays
that involved collisions with magnetic field structures. Much work followed in
which this process was modelled by particles impacting repeatedly (and
elastically)
against oscillating
heavy objects. The idea also found a natural application to particle
accelerators where imperfections in the confining plasma ring could be modelled
as periodic `kicks' in a very similar manner. Because of these applications,
and the simple form that the mappings take, they have become standard
problems in Hamiltonian systems with two degrees of freedom.
Some work has also been done for inelastic systems \cite{l90,e86}.

Several different accelerator models have been studied. Two of the most
important are the Fermi model and the Pustyl'nikov model.

\subsubsection{The Fermi models}

Ulam et al. \cite{u61} performed the first study of the
Fermi mechanism. The model
they used was
a particle bouncing elastically with constant speed between two walls --- one
fixed and one moving periodically. The impact maps arising
from such systems are known as {\em Fermi maps}.
They were primarily interested in the long-term behaviour of the particle and
the maximum velocity that could be attained.
Further studies can be found in \cite{b71,ll72,llc80,ll}.

Typically the grazing discontinuity set lies far below
the lowest bounding invariant curve which is why the chaotic/regular
trajectories
described in \S~\ref{sec:barriers} have not been observed for this
model.
A simplification that has been frequently made is to treat the position
of the oscillating wall as being fixed while allowing its velocity to
oscillate.
In this way, the time of the next impact can be calculated explicitly for
many wall motions and so the mapping itself can be written explicitly.
This is a good
approximation for high-velocity motions although at low velociies
it precludes the possibility of grazing.
For a sinusoidal wall motion the simplified
mapping is
\bq
\begin{array}{rll}
 u_{n+1}& =& \left| u_{n}+\sin(\psi_{n}) \right| \\
\psi_{n+1}& =& (\psi_{n}+A/u_{n+1}) \bmod 2\pi
\end{array}
\label{eq:fermi} \eq
where $A$ is the system parameter.
Note that the time between impacts is inversely proportional to the
velocity.

At low velocities there is a stochastic sea interspersed with
elliptic islands.
At higher velocities it has been proved (see \cite{llc80}) that for
wall velocities that are
$C^{3+\epsilon}$ there are invariant curves which cross the
phase space and act as upper
bounds to the velocity of the particle.
If the wall motion is given by, for example, a saw-tooth function then these
bounding curves disappear and the motion can become unbounded.
Indeed much of the subsequent work on this system was concerned with the
existence of such curves and the smoothness conditions of KAM theory.

\subsubsection{The Pustyl'nikov map}

Pustyl'nikov \cite{p78} examined a different system consisting of a ball
returning to an oscillating wall under the influence of gravity.
Again the position of the wall can be regarded as fixed for high
velocities and
this gives rise to the following simplified mapping

\bq
\begin{array}{rll}
u_{n+1}& =& \left| u_{n}+\sin(\psi_{n}) \right| \\
\psi_{n+1}& =& (\psi_{n}+Au_{n+1}) \bmod 2\pi
\end{array}
\label{eq:pust}\eq
for sinusoidal forcing.
Pustyl'nikov proved that even for analytic wall velocities parameters
can be chosen such that there is no bound to the velocity of the
particle. The difference between this case and the Fermi maps is due
to the time between impacts at high energies which now is proportional to the
velocity.

\subsubsection{The impact oscillator}
\label{sec:movwal}
In (\ref{eq:imposc}) the wall is stationary and the
forcing is on the spring. However it is an easy exercise to show that,
upon making the substitution $y=x-\gamma\cos(\omega t)$,
equation (\ref{eq:imposc}) becomes
\bq \begin{array}{rcll} \label{eq:movwal}
\ddot{y}+y & = & 0, \quad & y< \sigma- \gamma\cos(\omega t)  \\
\dot{y} & \mapsto & -r\dot{y}+(1+r)\gamma\omega\sin(\omega t), \quad &
y=\sigma-\gamma \cos(\omega t)
\end{array} \eq
This is the equation of motion for an unforced mass moving on a linear
spring and impacting against a wall whose position at time $t$ is given by
$y=\sigma-\gamma\cos(\omega t)$. Therefore, except for $\omega=1$, the
forced oscillator problem defined by (\ref{eq:imposc}) is equivalent to a
moving wall problem. This equivalence also holds for more general
periodic forcing as long as there is no power at the frequency
$\omega=1$. This relationship between the two systems is highly
relevant
since in physical situations the driving oscillations
may act on the wall or on the
mass/spring component.

In the Fermi model the particle moves with constant speed between
impacts with the vibrating wall. In the Pustyl'nikov model the particle moves
freely under a constant acceleration (linear potential).
The next `natural' case to consider is
a particle moving in a quadratic potential which, by the above result, is
the impact oscillator (for $\omega \neq 1$). So the impact oscillator fits
very naturally into the family of accelerator models.

The impact oscillator is  similar to
the Fermi model with chaotic behaviour at low energies  and regular
behaviour at higher energies. But there are
important differences, especially at high velocities.
These are also due to the time between impacts.
For the Fermi model the time between impacts is ${\cal O}(1/v)$ for large
velocities while for the Pustyl'nikov model it is ${\cal O}(v)$. For the
impact oscillator the time between impacts tends to $\pi$ as $v$ increases
as can be seen from equations (\ref{eq:pos}) and (\ref{eq:vel}).
So at high velocities the period of the oscillation is almost independent of
the amplitude. This property is reminiscent of linear oscillators and for this
reason we expect resonance effects to be important. This is indeed the case
and the behaviour at high velocities is much more complicated than for
the Fermi or Pustyl'nikov models, as we show in \S\ref{sec:highv}.

\subsection{Discontinuities in impacting systems}

We briefly discuss the different kinds of discontinuities that
can appear, and have been studied,  in impact maps.

A billiard with a smooth boundary has a smooth impact map and
introducing a discontinuity into the $n^{\rm{ th}}$ derivative
of the curvature at some point of the boundary leads to an impact
map with discontinuous $(n-1)^{{\rm th}}$ derivative.
This situation has been much
studied, especially with respect to proving ergodic and mixing properties and
testing the smoothness conditions of KAM theory.
Similarly, if for the simplified Fermi map, the wall velocity is smooth,
then the impact map is also smooth and introducing discontinuities into
the wall velocity or
a higher derivative of the wall velocity results in
a corresponding discontinuity
in the impact map.

When the map itself
is discontinuous, nearby trajectories that
straddle the discontinuity are separated.
When the discontinuity is in the first derivative or higher,
nearby trajectories
remain close together. For examples of such discontinuous systems, see
\cite{b71,hw83,ks89,lm86,wgc90}.

The grazing discontinuity is of a
different form. Nearby trajectories that straddle \SC\ are not separated.
Instead, one trajectory has an extra (low velocity)
impact and the distance between the
trajectories
is stretched as was described in the previous section.
It is perhaps helpful to think of a graze as being less severe
than a discontinuity in the mapping but more severe than a discontinuity in
the first derivative (if the extra low velocity impact is ignored then the
impact map is H\"{o}lder continuous with exponent $\frac{1}{2}$).
Grazing discontinuities are a natural feature of impacting systems, much
more natural than a discontinuity in the $n^{th}$ derivative, but have been
less thoroughly studied.

\section{Behaviour of the impact oscillator at high velocities}
\label{sec:highv}

In this section we examine the impact oscillator at high velocities in the
region above \SC\
where \PI\ is smooth.

We showed in \S\ref{sec:discont} that the elastic impact oscillator fits very
naturally into the family of accelerator models that includes the Fermi maps
and the Pustyl'nikov maps. This link motivates us to ask the following
question --- under what conditions is it possible for the velocity of the
mass to become unbounded?
To answer this it is
necessary to find conditions for the existence or non-existence of
bounding invariant curves.

Similar questions have also been asked of smooth systems.
The boundedness or otherwise of a particle
moving in a
one-dimensional smooth time-periodic potential
has been investigated by several authors, see Norris
\cite{n92} and references therein. Norris gave a sufficient condition for the
existence of bounding curves for a system of the
form
\bq\ddot{x}+g(x)=p(t) \label{eq:norris} \eq
where $p(t)$ is a sufficiently smooth periodic function, $g(x) \rightarrow
\infty \mbox{ as } x \rightarrow \pm \infty$ and either
\[ g(x)/x \rightarrow \infty \mbox { as } x \rightarrow \pm\infty, \quad
\mbox{ or }\]
\[ g(x)/x \rightarrow 0 \mbox{ as } x \rightarrow \pm \infty \]
The method of proof relies on the construction of a twist map and
breaks down for
functions $g(x)$ which  are `too close' to the linear oscillator. The obvious
example is the linear oscillator itself $g(x)=x$ for which (\ref{eq:norris})
has unbounded solutions whenever $p(t)$ has a
Fourier component with the resonant frequency 1.

The impact oscillator retains some of the characteristics of the linear
oscillator which describes its motion between impacts,
and for high velocity motions the time between
impacts tends to a constant. Therefore, the
frequency is approximately independent of the amplitude of the motion and we
can expect there to be resonant forcing frequencies.

We first consider the response of the impact oscillator
to sinusoidal forcing and show that it depends not just on the forcing
frequency but also on the position of the wall.
We use the impact map \PI\ to examine (\ref{eq:imposc}) at high
energies and give sufficient conditions on the parameter values
for the velocities to be
bounded for all time.
We also find parameter values for which there exist trajectories whose
velocity can become unbounded.
We then briefly look at the case of more general periodic forcing.

We start by considering the approximate behaviour of the system at large
velocities.

\begin{lemma}
For $\omega \neq 1 \mbox{ and } v \gg \max(1,\omega \gamma)$
the impact map has the following form
\begin{eqnarray}
v_{1}&=& \displaystyle v_{0} + f(\phi_{0}) + {\cal O}(\frac{1}{v_{0}})
\nonumber \\
\phi_{1}&=& \displaystyle (\phi_{0} + \alpha + \frac{g(\phi_{0})}{v_{0}}
+ {\cal O}(\frac{1}{v_{0}^{2}})) \bmod 2\pi
\label{eq:approx} \end{eqnarray}
where $\alpha$ is a constant independent of $(\phi_{0},v_{0})$.
\end{lemma}

{\bf Proof}
We assume that $v_{0}$ is large and,  in particular, that
the point $(\phi_{0},v_{0})$
lies above the set \SC\ which
is bounded. From
equation (\ref{eq:transc}), $t_{1}$ satisfies
\bq \label{eq:timp}
\sigma =(\sigma-\gamma \cos\omega t_{0})\cos(t_{1}-t_{0})-(v_{0}-\omega \gamma
\sin \omega t_{0}) \sin(t_{1}-t_{0})+\gamma \cos \omega t_{1}
\eq
and for large $v_{0}$
the time between impacts is close to $\pi$. So we let
$t_{1}-t_{0}=\pi+\delta$ where $\delta$ is small
and expanding (\ref{eq:timp}) in powers of $1/v_{0}$ gives
\[
\delta = \frac{2\sigma-\gamma(\cos \omega t_{0}-\cos
\omega(t_{0}+\pi))}{v_{0}}+{\cal O}(\frac{1}{v^{2}_{0}})
\]
where $g(t)$ as defined in the statement of the Lemma is
\bq g(t)=2\sigma-\gamma(\cos \omega t + \cos \omega (t+\pi))
\label{eq:g} \eq
Substituting this value for $t_{1}$ into equation (\ref{eq:vel}) we obtain
the following expression for $v_{1}$
\begin{eqnarray}
v_{1}&=&v_{0}
-\omega\gamma(\sin\omega t_{0}+
\sin \omega(t_{0}+\pi))  \nonumber \\
&& \mbox{} + \frac{g(t_{0})}{v_{0}}[\sigma-\gamma\cos\omega t_{0}
+(1-\gamma)\cos \omega(t_{0}+\pi)]
+{\cal O}(\frac{1}{v_{0}^{2}})
\end{eqnarray}
We let
\bq \label{eq:f} f(t)=-\omega\gamma(\sin\omega t + \sin \omega(t+\pi)) \eq
Replacing $t$ by the phase $\phi$ in (\ref{eq:g}) and (\ref{eq:f}) we
obtain
\begin{eqnarray}
v_{1}&=&v_{0}-\omega\gamma(\sin(\phi_{0})+\sin(\phi_{0}+\alpha))+{\cal O}
(\frac{1}{v_{0}}) \nonumber  \\
\phi_{1}&=&[\phi_{0}+ \alpha + \nonumber \\
&& \frac{2\omega\sigma-\omega\gamma(\cos(\phi_{0})+\cos(\phi_{0}+
\alpha))}{v_{0}} + {\cal O}(\frac{1}{v_{0}^{2}}) ] \bmod 2\pi
\label{eq:highv} \end{eqnarray}
where $\alpha = \omega\pi \bmod 2\pi$.
$\hfill{\Box}$

We now prove the following result.

\begin{theorem}
If $\omega \neq 2n, n \in \Bbb N^{+} \mbox{ and }
\sigma \neq 0$ then
the velocity of the impact oscillator is bounded for all time for all initial
conditions.
\label{th:bound} \end{theorem}

{\bf Proof}
We prove this by showing that at high velocities the impact
map is a perturbation of an integrable twist map of the form
\begin{eqnarray}
r_{1}&=&r_{0} \nonumber \\
\psi_{1}&=&\psi_{0}+ \beta(r)
\label{eq:moser} \end{eqnarray}
where $\beta ' \neq 0$.
Then, if certain other conditions are satisfied, Moser's small twist theorem
\cite{m62} guarantees the existence of infinitely many bounding invariant
curves.
First we consider the case $\omega \neq 1$.

We shall need the following Lemma.

\begin{lemma}
\label{lem}
Let $n\alpha\neq 0 \bmod 2\pi \mbox{ where }
n\in{\Bbb N}^{+}$. Then the functional equations
\begin{eqnarray}
F(\phi + \alpha)-F(\phi)&=&\sin(n\phi+n\alpha)+\sin(n\phi) \\
G(\phi + \alpha)-G(\phi)&=&\cos(n\phi+n\alpha)+\cos(n\phi)
\end{eqnarray}
are respectively solved by
\begin{eqnarray}
F(\phi)&=&-\cot(\frac{n\alpha}{2})\cos(n\phi) \\
G(\phi)&=&\cot(\frac{n\alpha}{2})\sin(n\phi)
\end{eqnarray}
\end{lemma}

{\bf Proof} The result is easily established by using the Fourier transform.
$\hfill{\Box}$

We now assume $v\gg\omega\gamma\cot(\alpha/2)$ and make
the following coordinate change
\begin{eqnarray}
w&=&v+\omega\gamma\cot(\frac{\alpha}{2})\cos(\phi) \nonumber \\
\psi&=&\phi-\frac{\omega\gamma}{v}\cot(\frac{\alpha}{2})\sin(\phi)
\end{eqnarray}
which, using the results of Lemma \ref{lem} for $n=1$,
transforms (\ref{eq:highv}) into
\begin{eqnarray}
w_{1}&=&w_{0}+{\cal O}(\frac{1}{w_{0}^{2}}) \nonumber \\
\psi_{1}&=&\psi_{0}+\alpha+\frac{2\sigma\omega}{w_{0}}+{\cal O}(\frac{1}
{w_{0}^{2}}) \label{eq:twist}
\end{eqnarray}
Physically, this coordinate transformation corresponds to finding the
approximate form of the bounding invariant curves, the existence of which
we now prove.

Now, following \cite{m62} we introduce a small parameter $\rho$ via the
transformation $w=\rho^{-1}r \mbox{ where } 0 < \rho \leq 1 \mbox{ and }
1 \leq r \leq 2$. Equation (\ref{eq:twist}) becomes
\begin{eqnarray}
r_{1}&=&r_{0}+{\cal O}(\rho^{3}) \nonumber \\
\psi_{1}&=&\psi_{0}+\alpha+\frac{2\sigma\omega\rho}{r_{0}}+{\cal O}(\rho^{2})
\label{eq:twist2} \end{eqnarray}

This is a
small perturbation of a mapping of the form (\ref{eq:moser}) and
we are now almost in a position to use Moser's small twist theorem.
First we note that the map (\ref{eq:twist2}) is analytic and so easily
satisfies the smoothness assumption of Moser's theorem.
Secondly, if $\sigma \neq 0$, then the twist condition $\beta ' \neq 0$ in
equation (\ref{eq:moser}) is also satisfied.
It only remains to demonstrate the curve intersection property for
(\ref{eq:twist2}), namely that any bounding
curve $C$ intersects its image $C'$. This follows from the area-preserving
property of \PI\ via the coordinate transformations we have made
since. Even though at low velocities the phase space is cut
and stretched by the set \SC, the area under a bounding
curve $C$ which does
not intersect \SC\ must be preserved.

The result  for $\omega \neq 1$
now follows by a straightforward application of Moser's small twist
theorem which guarantees the existence of invariant bounding
curves for (\ref{eq:twist}) for arbitrarily small $\rho$, which corresponds to
arbitrarily high velocities of the original system. Therefore the velocity of
the mass is bounded for all time.

Finally we consider the case $\omega = 1$.
This is a special case because the equations of motion
(\ref{eq:pos}) and (\ref{eq:vel}) do not apply. Instead the
motion after an impact at $(t_{0},v_{0})$ is given by
\begin{eqnarray*}
x(t)&=&\sigma\cos(t-t_{0})-(v_{0}+\frac{1}{2}\sin(t_{0}))\sin(t-t_{0})
\\
&& \mbox{} + \frac{1}{2}(t-t_{0})\sin(t) \label{eq:pos1} \\
v(t)&=&-\sigma\sin(t-t_{0})-(v_{0}+\frac{1}{2}\sin(t_{0}))\cos(t-t_{0})
\\
&& \mbox{} + \frac{1}{2}(t-t_{0})\cos(t)+\frac{1}{2}\sin(t) \label{eq:vel1}
\end{eqnarray*}
Proceeding as before we find that for $v \gg 1$ the impact map has the form
\begin{eqnarray}
v_{1}&=&v_{0}-\frac{\pi}{2}\cos(\phi_{0})+{\cal O}(\frac{1}{v_{0}})
\nonumber \\
\phi_{1}&=&\phi_{0}+\pi+\frac{1}{v_{0}}(2\sigma+\frac{\pi}{2}\sin(\phi_{0}))
+{\cal O}(\frac{1}{v_{0}^{2}}) \bmod 2\pi \label{eq:highv1}
\end{eqnarray}
If we now consider $\PI^{2}$ then (\ref{eq:highv1}) becomes
\begin{eqnarray}
v_{2}&=&v_{0}+{\cal O}(\frac{1}{v_{0}}) \nonumber \\
\phi_{2}&=&\phi_{0}+\frac{4\sigma}{v_{0}}+{\cal O}(\frac{1}{v_{0}^{2}})
\end{eqnarray}
and, for $\sigma \neq 0$, we can apply Moser's invariant
curve theorem as before.
This completes the proof of the thoerem. $\hfill{\Box}$

\subsection{Unbounded motion}

The resonant cases $\omega=2n$ do in
fact lead to unbounded motion for certain values of $\sigma$.
Since $\alpha=0$,
(\ref{eq:highv}) becomes
\begin{eqnarray}
v_{1}&=&v_{0}-2\omega\gamma\sin(\phi_{0})+{\cal O}(\frac{1}{v_{0}})\nonumber
\\
\phi_{1}&=&\phi_{0}+\frac{2\omega}{v_{0}}(\sigma-\gamma\cos(\phi_{0}))
+{\cal O}(\frac{1}{v_{0}^{2}}) \bmod 2\pi \label{eq:unbo}
\end{eqnarray}

We prove the following result

\begin{theorem}
For $\omega=2n, n \in \Bbb N^{+}$, there are unbounded trajectories if
$|\sigma| < |\gamma|$.
\end{theorem}

{\bf Proof}
Making the substitution $s=1/v$ into (\ref{eq:unbo}) we get
\begin{eqnarray}
s_{1}&=&s_{0}+2s_{0}^{2}\omega\gamma\sin(\phi_{0})+{\cal O}(s_{0}^{3})
\nonumber \\
\phi_{1}&=&\phi_{0}+2s_{0}\omega (\sigma-\gamma\cos(\phi_{0}))+
{\cal O}(s_{0}^{2}) \label{eq:mt}
\end{eqnarray}
We now define $\phi^{\ast} $ by
$\phi^{\ast} = \cos^{-1}(\sigma/\gamma)$ and
$\gamma\sin(\phi^{\ast}) < 0$
and let
$t_{i}=\phi_{i}-\phi^{\ast}$.

Assuming that $s,t \ll 1$ and calculating the first neglected term
in (\ref{eq:mt})
we obtain the following
\begin{eqnarray}
t_{1}&=&t_{0}(1 + 2\omega\gamma s_{0}\sin(\phi^{\ast}))+{\cal O}
(s_{0}^{2}t_{0},s_{0}t_{0}^{2}) \nonumber \\
s_{1}&=&s_{0}(1 + 2\omega\gamma s_{0}\sin(\phi^{\ast}))+{\cal O}
(s_{0}^{3},s_{0}^{2}t_{0})
\end{eqnarray}
By the definition of $\phi^{\ast}, 2\omega\gamma\sin(\phi^{\ast}) <0$
and so
for sufficiently small $s,t$,
\[ |s_{n}|<|s_{n-1}| \mbox{ and }
|t_{n}|<|t_{n-1}|  \]
and it is clear that $\lim_{n \rightarrow \infty} s_{n}=0$
which corresponds to a motion with unbounded velocity.
$\hfill{\Box}$

We can summarise the above results as follows. For the generic case
$\sigma \neq 0 \mbox{ and } \omega \neq 2n$ the motion is bounded at
sufficiently high velocities. When $\sigma=0$ the boundedness of the motion
is marginal and depends on the higher order terms that were
neglected in the above analysis. If the energy is in fact unbounded,
then the rate of energy gain is extremely slow.
The resonant frequencies are at $\omega=2n$
where unbounded motion occurs for $|\sigma| < |\gamma|$ and numerical
experiments indicate that this inequality is tight and for
$|\sigma| > |\gamma|$ the motion is bounded (in this case, the map
(\ref{eq:mt}) cannot be regarded as a small perturbation of an integrable map
of the form (\ref{eq:moser}) and Moser's theorem cannot be applied).

Thus the existence of bounding invariant curves depends crucially on
the position of the wall,
$\sigma$. For $|\sigma|$ sufficiently large (which means $\sigma \neq 0$ for
non-resonant $\omega$) the motion is bounded. In the mechanical
system described by (\ref{eq:imposc}) there are two ways of changing the value
of $\sigma$. One way is of course simply to move the wall. The other is to
add a constant term, $a$, to the forcing $\cos(\omega t)$.
Then a simple coordinate
change $x \rightarrow x-a$ recovers the initial
impact oscillator system with the new value of $\sigma$ given by
$\sigma \rightarrow \sigma -a$. This leads us to
the physically interesting, and rather counterintuitive, result that to
suppress
any unbounded behaviour it is enough just to add a constant force of sufficient
magnitude --- pointing either towards or away from the wall.

\subsection{General periodic forcing}

We can use the fact that the motion between impacts is linear to extend the
method to the case of more general periodic forcing functions. We
consider a forcing function $p(t)$ with period $2\pi/\omega$ and
Fourier components given by
\bq p(t)=\sum_{n=1}^{\infty}p_{n}\cos{n\omega t}+q_{n}\sin{n\omega t}
\label{eq:genforce} \eq
where, as described above, the constant term in the Fourier series has been
taken as zero and absorbed into the position of the wall. We now prove the
following result.

\begin{theorem}
Let $\sigma \neq 0$ and $p(t) \in C^{2+\epsilon}$. Then for a set of
irrational $\omega$ having full measure
the velocity of the impact oscillator is bounded for
all time for all initial conditions.
\end{theorem}

{\bf Proof}
In order to simplify the algebra we shall only consider an even periodic
forcing function
\bq p(t)= \sum_{n=1}^{\infty}p_{n}\cos(n\omega t) \eq
since the terms of the sine expansion are dealt with in an identical fashion.

The time of the next impact, $t_{1}$, is given by the first solution
of the equation
\begin{eqnarray}
\lefteqn{\sigma=(\sigma-\sum_{n=1}^{\infty}\gamma_{n} p_{n}\cos(n\omega t_{0}
))\cos(t_{1}-t_{0}) -} \nonumber \\
& &(v_{0}-\sum_{n=1}^{\infty}
n\omega\gamma_{n}p_{n}\sin(n\omega t_{0}))
\sin(t_{1}-t_{0})+
\sum_{n=1}^{\infty}\gamma_{n}p_{n}\cos(n\omega t_{1}) \label{eq:impfunc}
\end{eqnarray}
where $\gamma_{n} = 1/(1-n^{2}\omega^{2}) $.
If $p(t) \in C^{2+\epsilon}$ then the LHS of (\ref{eq:impfunc}) is
$C^{3+\epsilon}$
since $n\gamma_{n} \approx 1/n$ as $n \rightarrow \infty$.
So by the implicit function theorem $t_{1}(t_{0},v_{0}) \in C^{3+\epsilon}$.
Taking $v \gg 1$ and writing $t_{1}=t_{0}+
\pi + \delta$ we obtain the following equation for $\delta$
\begin{eqnarray} \label{eq:t1}
\sigma &=&(-\sigma+\sum_{n=1}^{\infty}\gamma_{n}p_{n}\cos(n\omega t_{0}))
\cos(\delta)+(v_{0}
-\sum_{n=1}^{\infty}n\omega\gamma_{n}p_{n}\sin(n\omega t_{0})
)\sin(\delta)  \nonumber \\
& & \mbox{} + \sum_{n=1}^{\infty}\gamma_{n}p_{n}\cos(n\omega(t_{0}+\pi+\delta))
\end{eqnarray}

Proceeding as before, we get the following approximate mapping at
sufficiently high
velocities
\begin{eqnarray}
v_{1}&=&v_{0}-\sum_{n=1}^{\infty}n\omega\gamma_{n}p_{n}(\sin n\phi_{0}+\sin
(n\phi_{0}+n\alpha))+{\cal O}(\frac{1}{v_{0}}) \nonumber \\
\phi_{1}&=&\phi_{0}+\alpha \nonumber \\
&& \mbox{} +\frac{2\sigma\omega-\sum_{n=1}^{\infty}\omega
\gamma_{n}p_{n}(\cos n\phi_{0}+\cos(n\phi_{0}+n\alpha))}{v_{0}}+{\cal O}
(\frac{1}{v_{0}^{2}})
\end{eqnarray}
where $\alpha = \omega \pi \bmod 2\pi$ and the mapping is $C^{3+\epsilon}$.
Once again, for $\alpha \neq 0$, Lemma \ref{lem} suggests the correct
coordinate change to reduce
the mapping to a sufficiently small perturbation of an integrable
twist map. This
change is
\begin{eqnarray} \label{eq:transgen}
w&=&v+\sum_{n=1}^{\infty}n\omega\gamma_{n}p_{n}\cot(\frac{n\alpha}{2})\cos
(n\phi) \nonumber \\
\psi&=&\phi-\frac{1}{v_{0}}\sum_{n=1}^{\infty}\omega\gamma_{n}p_{n}\cot
(\frac{n\alpha}{2})\sin(n\phi)
\end{eqnarray}
and is only well-defined if both of the Fourier series
in (\ref{eq:transgen}) converge. Manipulating the
coefficients of the first series slightly (if this series converges then so
does the second one), we get
\begin{eqnarray*}
\sum_{n=1}^{\infty} \left| n\omega\gamma_{n}p_{n}\cot(\frac{n\alpha}{2})
\right| &\leq & \sum_{n=1}^{\infty} \left|
\frac{n\omega \gamma_{n}
p_{n}}{\sin(\frac{n\alpha}{2})} \right| \\
&=&\sum_{n=1}^{\infty}\left|\frac{2n\omega \gamma_{n}p_{n}}{1-e^{in\alpha}}
\right| \end{eqnarray*}
This last sum is the classic `small-divisor' problem
and for $p \in C^{2+\epsilon}$, convergence of the series is guaranteed if
$\omega$ is irrational and has a continued fraction expansion that satisfies
certain conditions. The set of such $\omega$ has full measure.
Our smoothness condition on $p(t)$ was chosen so that the
perturbation is $C^{3+\epsilon}$ and we can now apply Moser's twist
theorem
exactly as for the sinusoidal case.
$\hfill{\Box}$

So for a set of forcing frequencies that has full measure, the motion
will be bounded for all $C^{2+\epsilon}$ forcing functions (for
$\sigma \neq 0$). It is not clear whether, for general
periodic forcing, unbounded behaviour can still be removed by increasing
the magnitude of the clearance.

{}From \S~\ref{sec:movwal} the impact oscillator with forcing
\[ p(t)=\sum_{n=1}^{\infty}p_{n}\cos(n\omega t) \]
is equivalent to the moving wall problem with wall motion given by
\[y(t)=\sum_{n=1}^{\infty}\gamma_{n} p_{n}\cos(n \omega t) \]
So we have the immediate corollary that for the
moving wall system the $C^{2+\epsilon}$ smoothness condition must be
replaced by a
$C^{4+\epsilon}$ condition on $y(t)$. This corresponds to a wall
velocity which is $C^{3+\epsilon}$.

\vspace{0.5in}

{\bf Acknowledgements} The author wishes to thank Chris
Budd for his invaluable support and advice and Felix Dux for many stimulating
discussions. This work was jointly funded by the Science and Engineering
Research Council and AEA Technology .

\newpage


\newpage
\begin{figure}
\caption{The mechanical impact oscillator.}
\label{fig:imposc}
\end{figure}
\begin{figure}
\vspace{0.5in}
\caption{Trajectories in the neighbourhood of a graze.}
\label{fig:graze}
\end{figure}
\begin{figure}
\vspace{0.5in}
\caption{Several trajectories of the impact oscillator
showing the phase space of \PI\ for
parameter values $\sigma=0.1$, \, $\omega=2.5$. The thick line is the
discontinuity set.}
\label{fig:pspace}
\end{figure}
\begin{figure}
\vspace{0.5in}
\caption{A {\em single} trajectory showing both regular and chaotic behaviour
for $\sigma=0, \, \omega=2.8$. The thick line is the discontinuity set.}
\label{fig:mixed}
\end{figure}
\begin{figure}
\vspace{0.5in}
\caption{The discontinuity set \SC\ for $\sigma=0, \,
\omega=5.3$.}
\label{fig:sc}
\end{figure}
\begin{figure}
\vspace{0.5in}
\caption{The dynamics close to $S$.}
\label{fig:singset}
\end{figure}
\begin{figure}
\vspace{0.5in}
\caption{Billiard coordinates}
\label{fig:bill}
\end{figure}
\begin{figure}
\vspace{0.5in}
\caption{Close-up of the partial barrier together with the last few iterates
of a trajectory on an elliptic curve before it reenters the stochastic sea.}
\label{fig:blowup}
\end{figure}
\begin{figure}
\vspace{0.5in}
\caption{An invariant curve, denoted by the solid line,
separating pieces of the partial barrier that allow
trajectories up and down.}
\label{fig:turn}
\end{figure}
\begin{figure}
\vspace{0.5in}
\caption{The phase space for $\sigma=0, \, \omega=2.85$. The discontinuity
set is also shown.}
\label{fig:case2}
\end{figure}
\end{document}